\documentstyle[prb,aps,epsf,twocolumn]{revtex}

\begin{document} 
\title{Magnetic anisotropy, first-order-like metamagnetic transitions and
large negative magnetoresistance in the single crystal of 
Gd$_{2}$PdSi$_3$}
 
\author{S.R. Saha, H. Sugawara, T.D. Matsuda, H. Sato, R. Mallik$^{@}$ and
E.V. Sampathkumaran$^{@}$}
 
\address{Department of physics, Tokyo Metropolitan University,
Hachioji-Shi, Tokyo 192-0397, JAPAN}
 
\address{$^{@}$Tata Institute of Fundamental Research, Homi
Bhabha Road, Colaba, Mumbai - 400005, INDIA} 
\date{\today}  
\maketitle 

\begin{abstract}
Electrical resistivity ($\rho$), magnetoresistance (MR),
magnetization, thermopower and Hall effect measurements on the single
crystal Gd$_{2}$PdSi$_3$, crystallizing in an AlB$_2$-derived
hexagonal structure are reported.  The well-defined minimum in $\rho$ at a
temperature above N\'eel temperature (T$_N$= 21 K) and large negative MR
below $\sim$ 3T$_N$, reported earlier for the polycrystals, are
reproducible even in single crystals. Such features are generally
uncharacteristic of Gd alloys. In addition, we also found interesting
features in other data, e.g., two-step first-order-like metamagnetic
transitions for the magnetic field along [0001] direction. The alloy
exhibits  anisotropy in all these properties, though Gd is a S-state ion.

\end{abstract} 
\vskip 0.5cm 
The observation of large negative magnetoresistance (MR) above respective
magnetic transition temperatures in some polycrystalline Gd
(also Tb, Dy) alloys is of considerable interest.\cite{1,2,3,4,5} Among
the Gd alloys, we have studied the transport properties of the compound,
Gd$_2$PdSi$_3$, crystallizing in an AlB$_2$-type structure, which has been found to show unusual nature. While this compound orders antiferromagnetically at
(T$_N$=) 21 K, there is unexpectedly a distinct minimum in the temperature
dependent electrical resistivity ($\rho$) at about 45 K. This minimum
disappears by the application of a magnetic field (H), thereby resulting
in large MR in the vicinity of T$_N$ [Ref. 3]. These properties are also
characteristic of Ce/U-based Kondo lattices, but uncharacteristic of Gd
systems, considering that the Gd-4f orbital is so deeply localised that it
cannot exhibit the Kondo effect. Though magnetic-polaronic effect (even in
metallic environments) has been  proposed in references 1-5 as one of
possible mechanisms behind this large MR,  its origin is not clear yet. It, however, appears that short-range correlation as a magnetic precursor
effect may be the primary ingredient\cite{5} for the origin of the
resistivity minimum above T$_N$ and negative MR. The importance of such
findings is obvious from similar recent reports from other
groups\cite{6,7,8,9,10} and among these the observation of $\rho$ minimum
and resultant colossal magnetoresistance (CMR) in a pyrochlore-based
oxide, Tl$_2$Mn$_2$O$_7$ [Ref. 7,8], has attracted recent attention. In
view of the importance of the observations on polycystals of this Gd
compound, we considered it important to confirm the findings on the single
crystals. With this primary motivation, we have investigated $\rho$, MR,
thermopower (S), Hall-effect and magnetization behavior on single crystals
of Gd$_2$PdSi$_3$, and these results are presented in this article.

Single crystals of Gd$_2$PdSi$_3$ have been prepared by the Czochralsky
pulling method using a tetra-arc furnace in an argon atmosphere. The
single-crystalline nature has been confirmed using back scattering x-ray
technique. The $\rho$, MR and Hall effect (employing a magnetic field of
15 kOe)  measurements have been performed by a conventional DC four-probe
method down to 1.2 K; the MR and Hall effect measurements have also been
performed as a function of H at 4.2 K. The magnetic measurements have been
carried out with a Quantum Design Superconducting Quantum Interference
Device. The thermopower data were taken by the differential method using
Au-Fe (0.07\%)-chromel thermocouples.

Fig. 1a shows the temperature dependence (1.2-300 K) of $\rho$ for the
sample with the current j//[$10\overline{1}0$] and j//[0001] in zero field. 
In Fig. 1b, the low temperature data, normalised to the 300 K value, in
the absence of a magnetic field as well as in the presence of 50 kOe (in
the longitudinal geometry, H//j) are shown.  In zero field, the $\rho$(T)
gradually decreases with decreasing temperature like in ordinary metals,
however, only down to about 45 K below which there is an upturn. There is
a kink at about 21 K for both directions, marking the onset of
magnetic ordering.\cite{3} The $\rho$, however, does not drop sharply at
T$_N$ expected due to the loss of spin-disorder contribution, but exhibits
a tendency to flatten or a  fall slowly with decreasing temperature.
Presumably, the magnetic structure could be a complex one, resulting in
the formation of superzone-zone boundary gaps in some portions of the
Fermi surface.  The application of a magnetic field, say H = 50 kOe, in the geometry discussed above, however depresses the $\rho$ minimum restoring metallic behavior in the entire temperature range of investigation. This naturally means 
that there is a large negative magnetoresistance, MR = $\Delta\rho/\rho=
[\rho(H)-\rho(0)]/\rho(0)$, at low temperatures, the magnitude of which
increases with decreasing temperature. A large negative value  close to -30\%  
could be seen for moderate fields (15 kOe) at 4.2 K (Fig. 2a), an indication 
of giant magnetoresistance.  Thus, all these features observed in polycrystals,
 are reproducible in single crystals as well.  It is obvious (Fig. 1a) that the 
absolute values of $\rho$ are relatively higher for j//[$10\overline{1}0$] than
 that for j//[0001]. It is to be noted that, though these values still fall in 
the metallic range, the temperature dependence is rather weak, for instance,
$\rho$(4.2K)/$\rho$(300K) is not less than 0.75 (in zero field), in sharp
contrast to a value of about 0.2 even for polycrystalline Lu$_2$PdSi$_3$
(Ref.  11).  It is not clear whether this fact is associated with some kind of
disorder effect on $\rho$ in magnetic sample compared to that in
nonmagnetic Lu$_2$PdSi$_3$ or with an intrinsic mechanism
responsible for the $\rho$ minimum.

\begin{figure}
\begin{center}
\epsfxsize=8.6cm \epsfbox{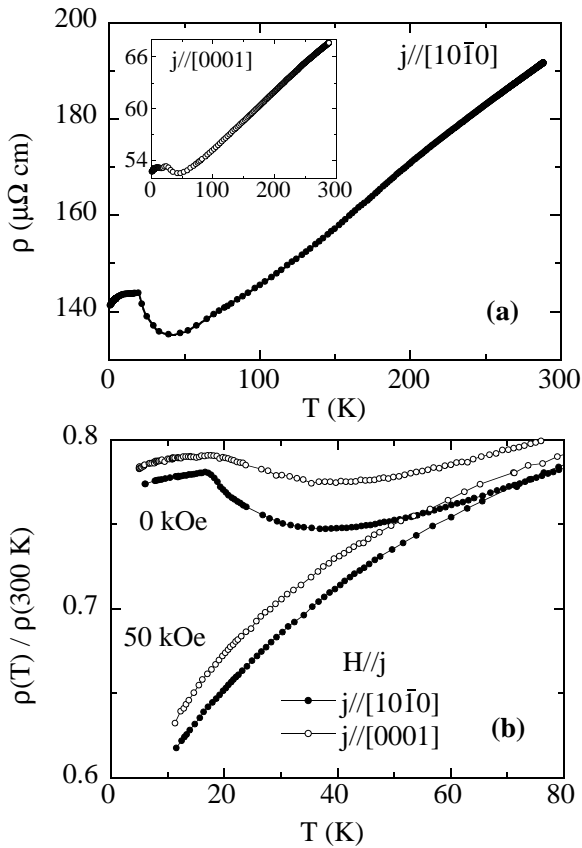}
\caption{(a) The electrical resistivity ($\rho$) of single crystalline 
Gd$_2$PdSi$_3$ as a function of temperature (1.2-300 K) for j//[$10\overline{1}0$] and 
j//[0001] (the inset). (b) The low temperature $\rho$ data in the absence of 
a magnetic field and in the presence of a field of 50 kOe is shown in an 
expanded form after normalizing to 300 K values.}
\end{center} 
\end{figure}

We have also measured MR as a function of H at 4.2 K with H varying from -15 kOe to 15 kOe, both in the longitudinal and transverse geometries for H//[0001] 
and H//[$10\overline{1}0$]. For H//[0001], the transverse MR (j//[$10\overline{1}0$]) is positive with a small magnitude, while the longitudinal MR (j//[0001]) is negative (see Fig. 2a).  The contribution from the anisotropic MR due to spin-orbit coupling is negligible for the Gd ion. The cyclotron contribution to the resistivity is also  small. Possible contribution resulting from the reduction of magnetic scattering due to the metamagnetic transitions should be the same for both geometries, since the field directions are the same. Therefore, the
anisotropy in MR reflects the anisotropy of the Fermi surface for the two
current directions. We believe that the conductivity parallel to [0001] is
favored by the disappearance of the magnetic superzone gaps in some
portions of the Fermi surface, resulting in a decrease of $\rho$ in the
longitudinal MR geometry for H//[0001]. However, it is interesting to note
that, for H//[$10\overline{1}0$], one sees negative MR for both geometries (j//[$10\overline{1}0$] and j//[0001]).  A noteworthy finding is that, for H//[0001], there are sharp changes in MR when measured as a function of H as indicated by
arrows in Fig. 2a, but occurring at different fields for the two
geometries due to the difference in the demagnetizing fields.  There is a
small hysteresis at the region around the sharp changes and we see similar
behavior even in the magnetization data (see below); such sharp variations
are absent for H//[$10\overline{1}0$]. All these results bring out anisotropic 
nature of MR.

\begin{figure}
\begin{center}
\epsfxsize=8.6cm \epsfbox{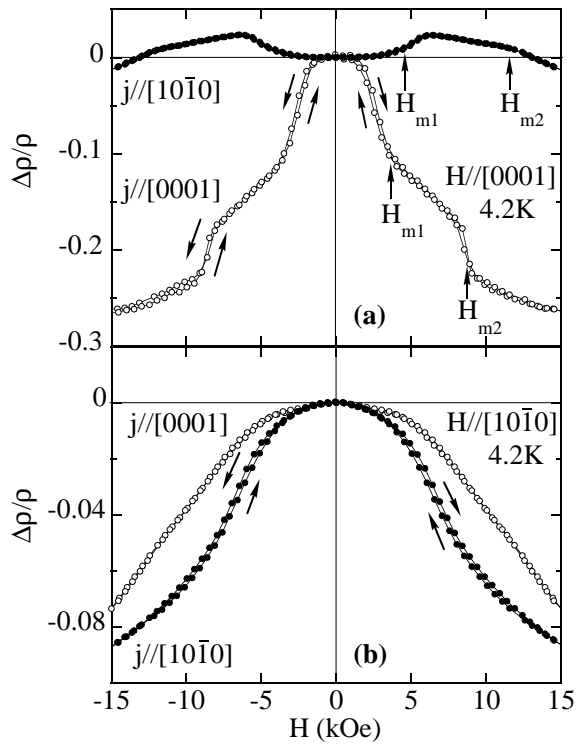} 
\caption{The magnetic field dependence of magnetoresistance for
Gd$_2$PdSi$_3$ in the transverse and longitudinal geometries, as labelled
in the figure. The arrows indicate the directions of the field sweep.}
\end{center} 
\end{figure}

Fig. 3a shows the temperature dependence of magnetic susceptibility
($\chi$), measured in the presence of a field of 1 kOe for both H//[0001]
and H//[$10\overline{1}0$]. There is a well-defined peak in $\chi$ at 21 K confirming the antiferromagnetic nature of the magnetic transition; below 21 K,
however, there is only a small difference in the values for these two
geometries. The paramagnetic Curie-temperature turns out to be the same for
both geometries, with the same magnitude as that of T$_N$, however,
with a positive sign suggesting the existence of strong ferromagneticcorrelations. There is no difference between

\begin{figure}
\begin{center}
\epsfxsize=10.5cm \epsfbox{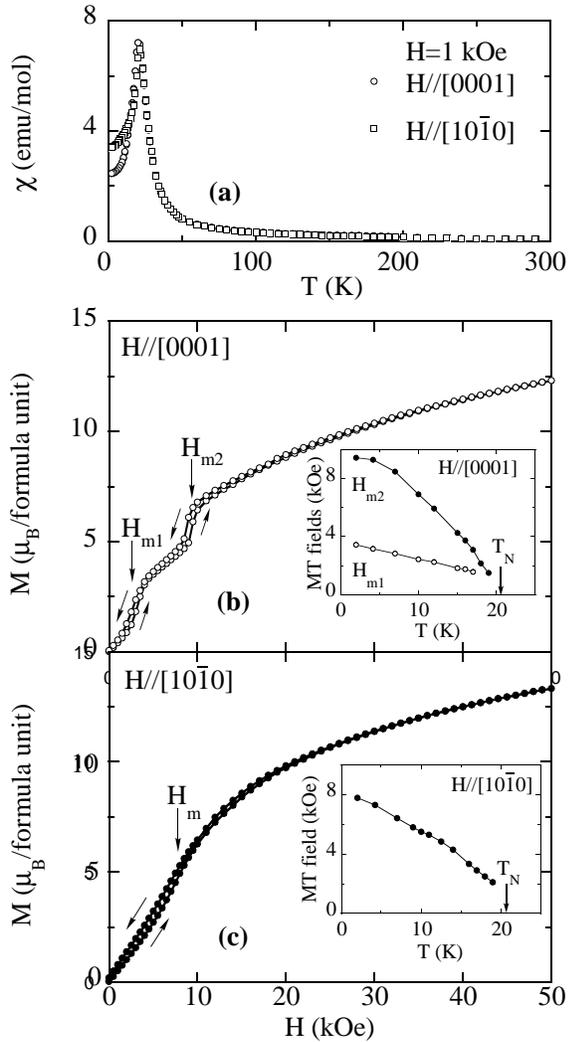} 
\caption{(a) The magnetic susceptibility versus temperature (2-300 K) for
single crystals of Gd$_2$PdSi$_3$. The isothermal magnetization behavior
at 2 K for H//[0001] and H//[$10\overline{1}0$] are shown respectively in (b) and (c);
the insets show the metamagnetic transition (MT) fields obtained as
described in the text as a function of temperature.}
\end{center}
\end{figure}

field-cooled (FC) and
zero-field-coooled (ZFC) $\chi$ values below 21 K, unlike the situation in
polycrystals.\cite{3} This suggests that such difference in FC and ZFC
data in polycrystals is not intrinsic to this material. This fact supports
our earlier conclusion\cite{3} that this alloy is not a spin-glass, unlike
U$_2$PdSi$_3$ (Ref. 12).

The isothermal magnetization (M) behavior at 2 K is shown in Fig. 3b, both
for increasing and decreasing fields. For the field along [0001], there are
two step-like metamagnetic transitions, one around 3 kOe and the other
around 9 kOe. Apparently, there is a small hysteresis around these
transitions, indicating first-order nature of the transitions. The inset
of figure 3b shows the metamagnetic transition fields H$_{m1}$ and
H$_{m2}$ (the magnetic fields corresponding to the highest dM/dH at the
low-field and high-field transitions, respectively) versus temperature;
both H$_{m1}$ and H$_{m2}$ decrease with increasing temperature. M vs H
for H//[$10\overline{1}0$] also shows a faint meta-magnetic anomaly (Fig. 3c), however with M varying relatively smoothly with H, unlike the situation for
H//[0001]; the inset shows the characteristic magnetic field, H$_m$
(estimated in the same way as H$_{m1}$ and H$_{m2}$). The results
establish the existence of anisotropy in the isothermal magnetization. The
observation of metamagnetic transitions are consistent with the anomalies
in MR, discussed above.

\begin{figure}
\begin{center}
\epsfxsize=8.6cm\epsfbox{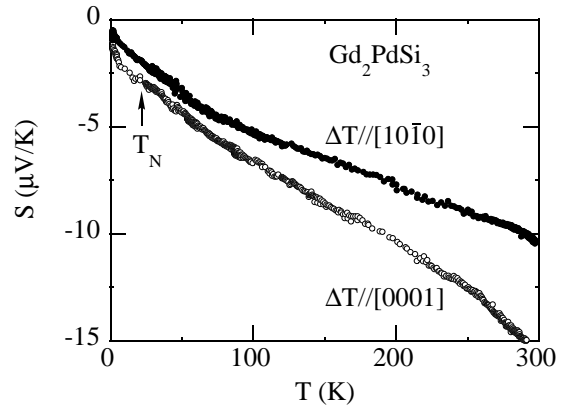}
\caption{The thermopower as a function of temperature for two different
directions of thermal gradient on Gd$_2$PdSi$_3$ single crystals.}
\end{center}
\end{figure}

Fig. 4 shows the temperature dependence of thermopower. The absolute value
is large at 300 K as in the case of Lu$_2$PdSi$_3$ (Ref. 3). Therefore,
the large S might arise from 4d band of Pd as in the case of 3d band of Co in 
YCo$_2$.\cite{13} There is no anomaly, however, at T$_N$. S decreases with
decreasing temperature and the features are qualitatively the same as
those observed in the non-magnetic Lu$_2$PdSi$_3$.\cite{3} Though the
overall S behavior mimics the one in polycrystals,\cite{3} there is a
distinct anisotropy in the values when measured along different
directions, that is, the value of S depends on whether the temperature
gradient, $\Delta$T, is parallel to [$10\overline{1}0$] or [0001]. The absence of any
peak-like behavior expected for Kondo system\cite{14} indicates the
absence of Kondo effect in  Gd$_2$PdSi$_3$.

\begin{figure}
\begin{center}
\epsfxsize=8.6cm \epsfbox{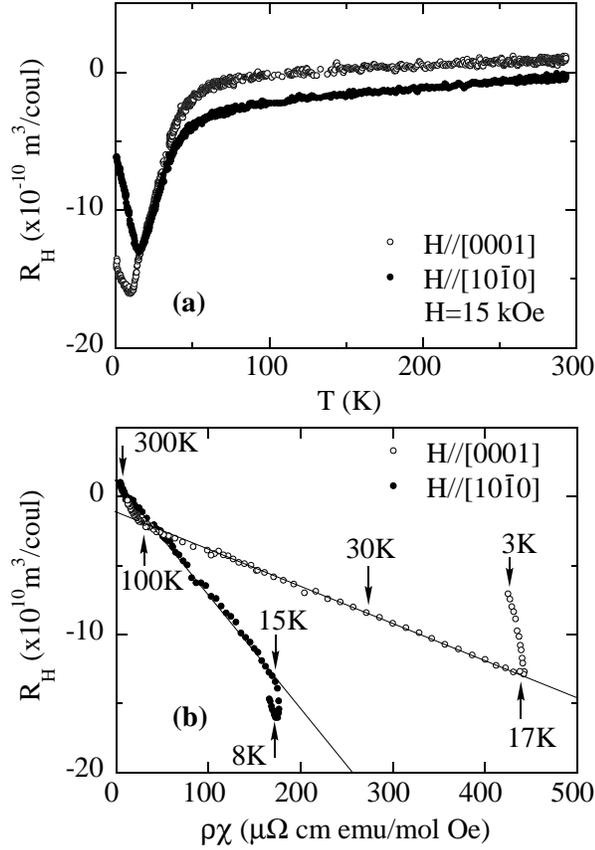} 
\caption{The temperature dependence of
Hall coefficient, R$_H$, for Gd$_2$PdSi$_3$ single crystals with two
different orientations is plotted in (a). In (b) the values of R$_H$ are
plotted as a function of the electrical resistivity times magnetic
susceptibility, with temperature as an intrinsic parameter. The
temperature region in which the R$_H$ varies linearly  is shown by drawing
continuous lines through the data points. For H//[0001], it is obvious that
there is a deviation from this line above 100 K, besides the one near
T$_N$.}
\end{center}
\end{figure}

The temperature dependence of Hall coefficient (R$_H$), shown in Fig. 5a,
also reflects anisotropic nature of this material. The R$_H$ shows large
temperature dependence, in contrast to the temperature independent
behavior in Lu$_2$PdSi$_3$ (Ref. 3), with a negative peak for both 
geometries, H//[0001] and H//[$10\overline{1}0$], in the vicinity of T$_N$, 
however at slightly different temperatures (the reason for which is not clear).
Clearly there is a dominant 4f contribution in the Gd case. The Hall
effect of magnetic metals like those of Gd is generally a sum of two terms
- an ordinary Hall effect (R$_0$) due to Lorentz force and an anomalous
part arising from magnetic scatterring (skew scatterring). Thus, in the
paramagnetic state, R$_H$= R$_0$ + A$\rho$$\chi$, where A is a constant.
Using this relation, R$_0$ is estimated by plotting R$_H$ versus
$\rho$$\chi$ (Fig. 5b). From Fig. 5b, it is obvious that the plot is
linear for H//[$10\overline{1}0$] in the paramagnetic state with a value of R$_0$=
0.92$\times$10$^{-10}$ m$^{3}$/coul.  However, for H//[0001], there is 
a distinct change in the magnitude as well as in the sign around 110 K 
as if there is a change in the sign of the carrier; 
(1.6$\times$10$^{-10}$ m$^{3}$/coul and -1.3$\times$10$^{-10}$ m$^{3}$/coul 
for T$>$ 110 K and T$<$110 K, respectively). Below 17 K, the data however 
deviate from the high temperature linear variation as the state is no 
longer paramagnetic.

\begin{figure}
\begin{center}
\epsfxsize=8.6cm \epsfbox{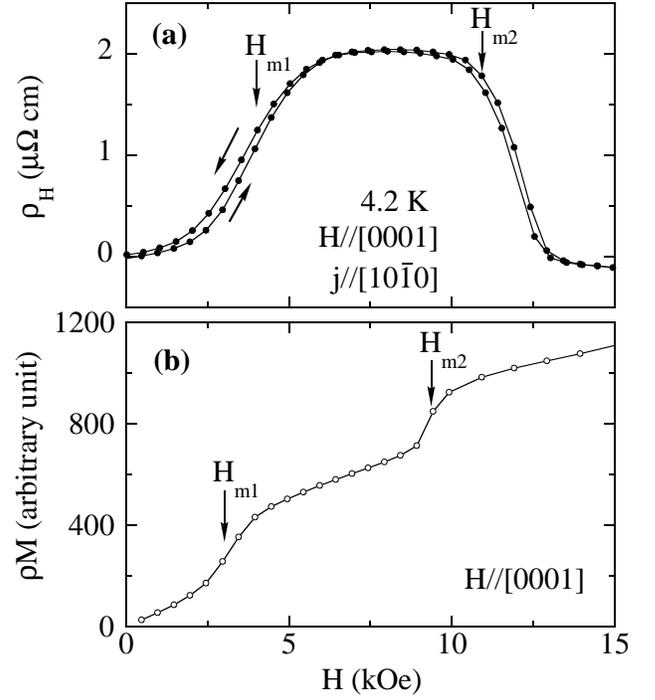}
\caption{(a) The field dependence of Hall resistivity in Gd$_2$PdSi$_3$
for H//[0001]. The arrows indicate the directions of the field sweep. (b)
Field dependence of $\rho$M for H//[0001].}
\end{center}
\end{figure} 

Fig. 6a shows the field dependence of Hall resistivity ($\rho$$_H$) for
H//[0001]. $\rho$$_H$ shows distinct anomaly across the two metamagnetic
transition fields and traces the hysteresis. Again considering the concept
of anomalous Hall effect discussed in the above paragraph, the field
dependence of the corresponding $\rho$M is ploted in Fig. 6b.
Theoretically, when anomalous Hall effect is dominant, $\rho$$_H$ should
vary linearly with $\rho$M. In other words, $\rho$$_H$ and $\rho$M
should vary in the same way with the corresponding applied fields. But in
the present case, the field dependence of $\rho$$_H$ completely differ from 
that of $\rho$M, particularly around H$_{m2}$.  This fact
strongly indicates the modification of the Fermi surface across the
metamagnetic anomaly.

Summarising, we have explored anisotropy in the
transport and magnetic properties on the single crystal Gd$_2$PdSi$_3$.  Possibly the anysotropic
exchange interaction due to crystalline anisotropy and the anisotropy in the 
Fermi surface are responsible for the observed anysotropy. Interestingly, there are 
magnetic field induced first-order-like magnetic transitions in the magnetically ordered state, resulting in large MR and consequent Fermi surface modification. Recently, first-order transition has been reported in another Gd-alloy,
Gd$_5$(Si,Ge)$_4$ (Ref. 15), which has been found to be a simultaneous 
crystallograpic and magnetic transition, and in view of this it is of interest 
to explore whether there is any structural transition with the application of H in our case as well.  In short, the single crystal of Gd$_2$PdSi$_3$  
exhibits interesting features. Above all, there is a well-defined $\rho$ minimum above T$_N$, the origin of which is still not completely clear; the
negative MR persists till about 3T$_N$ even in single crystals with the
magnitude gradually increasing with decreasing temperature towards T$_N$.
The results overall establish that this compound is a novel magnetic
material.

This work has been partially supported by a grant-in-Aid for Scientific
Research from the Minstry of Education, Science and Culture of Japan


\begin{references}
 
\bibitem{1}E.V. Sampathkumaran and I. Das, Phys. Rev. B {\bf51}, 8631
 (1995).
 
 \bibitem {2} R. Mallik, E.V. Sampathkumaran, P.L. Paulose, and V.
 Nagarajan, Phys. Rev. B {\bf 55}, R8650 (1997).
 
\bibitem {3} R. Mallik, E.V. Sampathkumaran, M. Strecker, and G. Wortmann,
Europhys. Lett., {\bf41,} 315 (1998); R. Mallik, E.V. Sampathkumaran, P.L.
Paulose, H. Sugawara, and H. Sato, Pramana - J. Phys. {\bf51}, 505 (1998).
 
\bibitem {4} R. Mallik, E.V. Sampathkumaran, and P.L. Paulose, Solid State
 Commun. {\bf106}, 169 (1998).
 
\bibitem {5} R. Mallik and E.V. Sampathkumaran, Phys. Rev. B {\bf58}, 9178
 (1998).
 
\bibitem {6}J.Y. Chan, S.M. Kauzlarich, P. Klavins, R.N. Shelton, and D.J.
Webb, Phys. Rev. B {\bf57}, R8103 (1998).
 
\bibitem{7}Y. Shimikawa et al, Nature (London) {\bf379}, 53 (1996).
 
\bibitem{8}M.A. Subramanian et al., Science {\bf273}, 81 (1996).
 
\bibitem{9}P. Majumdar and P.B. Littlewood, Phys. Rev. Lett. {\bf81}, 1314
(1997).
 
\bibitem{10}J.A. Alonso, J.L. Martinez, M.J. Martinez-Lope, M.T. Casals,
and M.T. Fernandez-Diaz, Phys. Rev. Lett. 1999.
 
\bibitem{11}There is a drop in $\rho$ for the Lu sample at 0.96 K with
decreasing temperature as if there is a superconducting transition. It is
at present not clear whether it is intrinsic.
 
\bibitem{12}D.X. Li, Y. Shiokawa, Y. Homma, A. Uesawa, A. Donni, T.
Suzuki, Y. Haga, E. Yamamoto, T. Honma, and Y. Onuki, Phys. Rev. B
{\bf57}, 7434 (1998).
 
\bibitem{13}H. Sugawara, T. Nishigaki, Y. Kobayashi, Y.  Aoki, and H.
Sato, Physica B{\bf230-232}, 179 (1997).
 
\bibitem{14}C. S. Garde and J. Ray,  Phys. Rev. B {\bf51}, 2960 (1995); J.
Sakurai, Transport and thermal properties of f-electron systems, edited by
G. Oomi , H. Fujii and T. Fujita (Plenum Press, New York) 1993, p. 165; A.
K.  Bhattacharjee and B. Coqblin, Phys. Rev. B{\bf13}, 3441 (1976).

\bibitem{15}L. Morellon, J. Stankiewicz, B. Garcia-Landa, P.A. Algarabel,
and M.R. Ibarra, App. Phys. Lett. {\bf73}, 3462 (1998); L. Morellon, P.A.
Algarabel, M.R. Ibarra, J. Blasco, B. Garcia-Landa, Z. Arnold, and F.
Albertini, Phys. Rev. B {\bf58}, R14721 (1998).
\end{references}
\end{document}